\documentclass[iop]{emulateapj}
\pdfoutput=1
\usepackage{natbib}
\usepackage{threeparttable}
\usepackage{amsmath}
\usepackage{amssymb}

\newcommand{\HI}{\ensuremath{\mbox{\rm \ion{H}{1}}}}

\renewcommand{\t}[1]{\mathrm{#1}}

\newcommand{\htwo}{\ensuremath{\mbox{H$_2$}}}

\newcommand{\zsun}{\ensuremath{\mathcal{Z}_\odot}}

\newcommand{\kms}{\mbox{km~s$^{-1}$}}

\newcommand{\av}{\ensuremath{\mbox{$A_{\rm V}$}}}

\shortauthors{Imara \& Loeb}

\begin{document}

\title{Limits on Intergalactic Dust During Reionization}

\author{Nia Imara \& Abraham Loeb}
\affil{Harvard-Smithsonian Center for Astrophysics, 60 Garden Street, Cambridge, MA 02138}

\email{nimara@cfa.harvard.edu}

\begin{abstract}
In this \emph{Letter}, we constrain the dust-to-gas ratio in the intergalactic medium (IGM) at high redshifts.  We employ models for dust in the local Universe to contrain the dust-to-gas ratio during the epoch of reionization at redshifts $z\sim 6-10$.  The observed level of reddening of high redshift galaxies implies that the IGM was enriched to an intergalactic dust-to-gas ratio of less than 3\% of the Milky Way value by a redshift of $z=10$.
\end{abstract}



\section{Introduction}
Dust in the intergalactic medium (IGM) must be accounted for in precision measurements of the galaxy luminosity function, the determination of accurate distances to standard candles, and the corresponding measurements of cosmological parameters (e.g., Riess et al. 2000; Goobar et al. 2002; Corasaniti 2006; Dijkstra \& Loeb 2009; M\'enard et al 2010a). Moreover, the presence of intergalactic dust may contaminate the cosmic microwave background and cosmic infrared background (e.g., Aguirre 1999).  From a theoretical point of view, given that dust plays an essential role in the evolution of stellar populations and galaxies, placing constraints on the dust content of the IGM can contrain the history of the Universe's first stars and galaxies during the epoch of reionization (Loeb \& Furlanetto 2013).

Outflows from early Population III and II stars, supernova explosions, and galactic winds are all mechanisms that may have contributed to the expulsion of dust from galaxies into the IGM.  Several studies have modeled the enrichment of the IGM by heavy elements, (e.g., Cen \& Ostriker 1999; Aguirre et al. 2001; Theuns et al. 2002; Furlanetto \& Loeb 2003; Dav\'e \& Oppenheimer 2007; Dav\'e et al. 2011).  The same outflows that carry metals out of galaxies also entrain dust.  Various studies have used observations of quasar colors to estimate the amount of intergalactic dust at intermediate redshifts ranging from $z\sim 0.5-1$,  (e.g., Wright \& Malkan 1987; M\"ortsell \& Goobar 2003; M\'enard et al. 2010b; Johansson \& M\"ortsell 2012).  

Despite recent advances, a number of questions about intergalactic dust remain unsettled.  For instance, is it possible to disentangle reddening due to the dust within the interstellar medium (ISM) of galaxies from reddening due to dust in the IGM?  Assuming that outflows from galaxies carry dust as well as metals, how far beyond their galaxy of origin can dust grains be transported?  Which fraction of the dust survives galactic outflows?  Aguirre (1999) and Bianchi \& Ferrara (2005) demonstrated that small dust grains may be disproportionately destroyed in outflows, and grains larger than $\sim 0.1~\mu \mathrm{m}$ may remain undetected in quasar surveys at optical wavelengths.

The outline of this \emph{Letter} is as follows.  In \S 2, we calculate the optical depth due to dust in the IGM as a function of redshift.    In \S 3, we show how the integrated optical depth depends on observed wavelength as well as the metallicity of the IGM.  In \S 3.2, we use the reddening observed for high-redshift galaxies to derive a new upper limit on the intergalactic optical depth at high redshifts. Finally, we discuss the implications of our work and summarize our conclusions in \S 4.  Throughout our discussion, we assume the standard values for the cosmological parameters: $H_0 = 67.3~\kms~\mathrm{Mpc}^{-1}$, $\Omega_m = 0.3$, and $\Omega_\Lambda =0.7$ (Planck Collaboration XI 2015).

\section{Model}
\subsection{Optical depth estimate}

The integrated optical depth due to dust, $\tau_d$, along the line-of-sight to a galaxy at redshift, $z_s$, is given by, 
\begin{equation}\label{eq:tau1}
\tau_d = \int n_d \sigma_{d,\lambda} \mathrm{d} l,
\end{equation}
where $n_d$ is the number density of dust particles, and $\sigma_{d,\lambda}$ is the averaged wavelength-dependent cross section per dust grain.  The cosmological path length, $\mathrm{d} l$, can be written as 
\begin{equation}
\mathrm{d} l = c~\mathrm{d} t = \frac{\mathrm{d} a}{\dot{a}} = \frac{\mathrm{d} a}{aH}, 
\end{equation}
where $c$ is the speed of light, $t$ is time, $a=(1+z)^{-1}$ is the scale factor, and $H$ is the Hubble parameter.  At redshifts of interest here,
\begin{equation}
H(z) = H_0 [\Omega_m (1+z)^3 + \Omega_\Lambda ]^{1/2},
\end{equation}
and since $H=\dot{a}/a = -\dot{z}/(1+z)$, we may rewrite Equation (1) as
\begin{equation}\label{eq:tau2}
\tau_d = c \int_0^{z_s}  n_d \sigma_d \frac{\mathrm{d} z}{H(1+z)},
\end{equation}
where $z_s$ is the source redshift.

Because little is known about the dust properties in the IGM (e.g., density, grain size, cross section), it is advantageous to express $n_d$ and $\sigma_{d,\lambda}$ in terms of a couple of free parameters.  Exploiting the fact that $\sigma_{d,\lambda} = m_d\kappa_\lambda$, where $m_d$ is the average mass of a dust grain and $\kappa_\lambda$ is the dust opacity, Equation (4) becomes
\begin{equation}\label{eq:tau}
\tau_d = c \int_0^{z_s}  n_p m_p \kappa_\lambda \frac{\mathrm{d} z}{H(1+z)}\times \mathrm{DGR},
\end{equation}
where $m_p$ is the proton mass and $n_p=2.5\times 10^{-7}(1+z)^3~\mathrm{cm}^{-3}$ is the mean proton number density.\footnote{We adopt the most recent value for $n_p$ from the WMAP Cosmological Parameters website, http://lambda.gsfc.nasa.gov/product/map/current/parameters.cfm.}  We define the dust-to-gas mass ratio, $\mathrm{DGR}\equiv n_d m_d/(n_p m_p)$, as an averaged quantity; namely, $n_d$ and $m_d$ represent the average number density of dust grains and the average mass of a dust grain along the line-of-sight.  Here, $\lambda$ in the subscripts of $\sigma_{d,\lambda}$ and $\kappa_\lambda$, refers to the rest frame wavelength along the line-of-sight.  Thus, keeping in mind that $\tau_d$ is the optical depth in the observer's reference frame, $\kappa_\lambda$ in Equation (5) depends on the observed wavelength via $\lambda=\lambda_\mathrm{obs}/(1+z)$.

\subsection{Dust models}
Our ignorance about the properties of IGM dust are encapsulated in the variables $\kappa_\lambda$ and DGR in Equation (5).  For reference, we consider models of dust in the local Universe to calibrate the properties of dust in the high-redshift Universe.  In particular, we use the models of Weingartner \& Draine (2001), Li \& Draine (2001), and Draine (2003), for the opacity of dust grains composed of a mixture of carbons and silicates and having a log-normal size distribution.\footnote{B. T. Draine's website, http://www.astro.princeton.edu/ $\sim$draine/dust/dustmix.html, provides tables of $\kappa_\lambda$ values for three different models corresponding to three different values of the selective extinction, $R_V$, which we use in Equation (\ref{eq:tau}).}  We study the sensitivity of our results to $R_V$, which depends on the grain composition and size distribution.

We also explore the consequences of varying the DGR, which correlates with metallicity, $\mathcal{Z}$.  R\'{e}my-Ruyer et al. (2014) evaluated the gas-to-dust ratio of nearby galaxies spanning the range between $1/50~\zsun$ and 2 \zsun, finding that the observed trend has a power-law slope of $\sim -3$ for metallicities\footnote{Metallicity is defined here as the abundance of oxygen with respect to hydrogen, $\mathcal{Z}=\mathrm{O/H}$.  Following Asplund et al. (2009), we assume $\mathrm{(O/H)}_\odot = 4.90 \times 10^{-4}$; i.e., $12 + \log(\mathrm{O/H})_\odot = 8.69$.}  below $\sim 1/4$ \zsun. The authors provide alternative functional forms for the gas-to-dust ratio versus metallicity relationship, depending on the CO-to-\htwo~conversion factor they used to estimate the total amount of molecular mass in a galaxy.  We adopt the function they derive assuming a metallicity-dependent conversion factor (as opposed to the standard Milky Way CO-to-\htwo~conversion factor).  In terms of the DGR, 
\begin{equation}\label{eq:dgr}
 \log\left(\frac{\mathrm{DGR}}{\mathrm{DGR}_\odot}\right) = 
  \begin{cases} 
   \log \left(\frac{\mathcal{Z}}{\zsun}\right)   &     \text{if } \mathcal{Z} > 0.26\zsun \\
   3.15\log \left(\frac{\mathcal{Z}}{\zsun}\right) + 1.25 & \text{if } \mathcal{Z} \le 0.26\zsun ,
  \end{cases}
\end{equation}
where $\log(\mathrm{DGR}_\odot)=-2.21$ (Zubko et al. 2004).

\begin{figure}
\epsscale{1.2}
\plotone{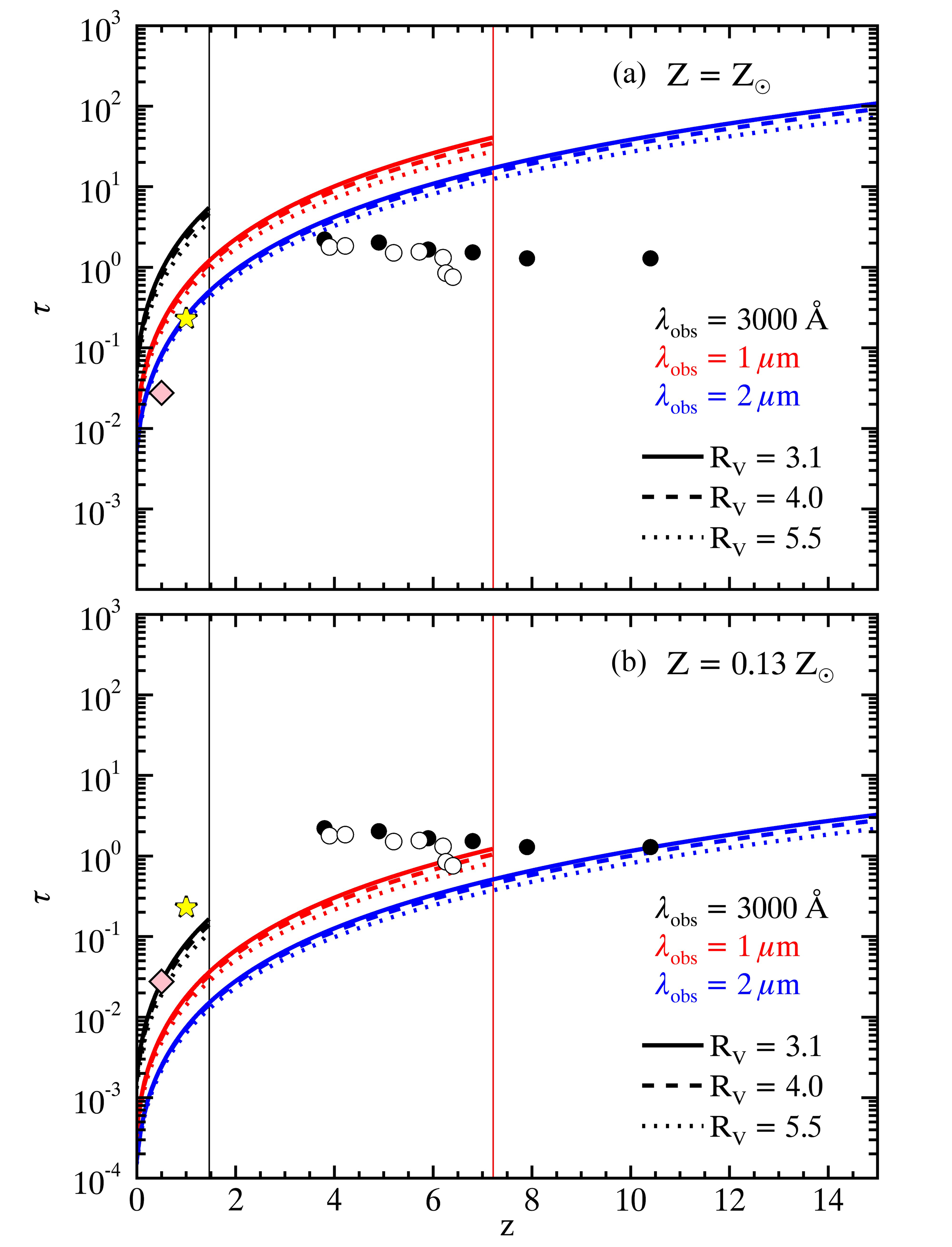} 
\caption{Extinction optical depth due to dust, $\tau_d$, as a function of redshift, $z$, for three different observed wavelengths, $\lambda_\mathrm{obs}=3000$ \AA, $1~\mu\mathrm{m}$, and $2~\mu\mathrm{m}$ (black, red, and blue, respectively).  The calculations assume a dust-to-gas ratio (DGR) corresponding to \textbf{(a)} \zsun~and \textbf{(b)} $0.13\zsun$.  Overplotted are data from M\'enard et al. (2010; diamond), Johansson \& M\"ortsell (2012; star),  Bouwens et al. (2015; filled circles), Gallerani et al. (2010; open circles).  For $\mathcal{Z}=\zsun$, by $z=15$, the maximum optical depths at $\lambda_\mathrm{obs}=2~\mu\mathrm{m}$ are $\tau_d=81.6$, 73.8, and 58.3, for $R_V=3.1$, 4.0, and 5.5. \label{fig1}}
\end{figure}

\begin{figure*}
\includegraphics[width=\textwidth]{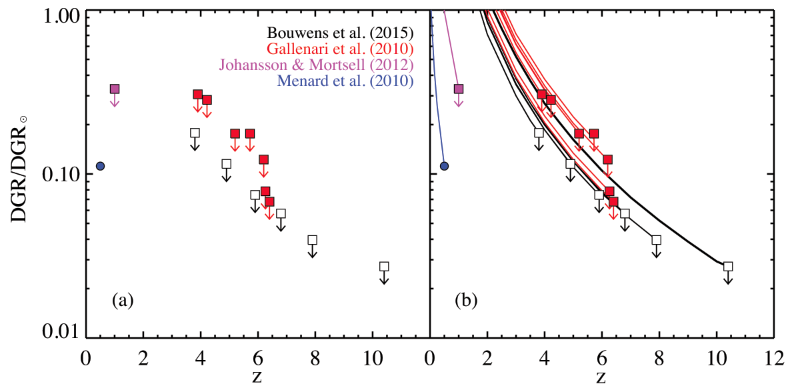} 
\caption{(a) Dust-to-gas ratio (DGR) as a function of $z$, derived from the extinction measurements cited in the legend.  The DGRs are expressed in units of $\mathrm{DGR}_\odot = 1/162$ (Zubko et al. 2004). All of the data points are upper limits, with the exception of the M\'enard et al. detection, labeled with a blue circle.  (b) Same as (a), with further constraints on the DGR at intermediate redshifts.  The curves make use of the fact that for any redshift along the line of sight to a source at $z_s$, there is an upper limit on the DGR based on the optical depth limit up to that redshift. \label{fig2}}
\end{figure*}

\section{Results}
\subsection{Limits on the Optical Depth of IGM Dust as a Function of Redshift}
Figure 1a shows the optical depth due to intergalactic dust as a function of redshift, calculated from Equations (5) and (6), for $\mathcal{Z}=\zsun$ and three observed wavelengths, $\lambda_\mathrm{obs} = 0.3$, 1, and 2 $\mu$m.  Photons emanating from the source will undergo Lyman-$\alpha$ absorption at $\lambda = 1216$ \AA.  For each $\lambda_\mathrm{obs}$, we indicate the redshift beyond which Lyman-$\alpha$ absorption obscures our view and for which the $\tau_d$-$z$ relation has little meaning, with a vertical line.  We discuss the overplotted data points in \S 3.2.
  
To illustrate the sensitivity of the results to dust properties, we also calculate the $\tau_d$-$z$ relationship for different values of $R_V$, including $R_V=3.1$ (solid lines), corresponding to the typical value through diffuse Milky Way clouds;  $R_V=4.0$ (dashed line); and $R_V=5.5$ (dotted line).  The selective extinction, defined as $R_V \equiv \av/(A_B - \av)$, defines the slope of the extinction curve in the optical band.  Smaller grains tend to scatter light at shorter wavelengths, yielding a steeper slope and yielding smaller values of $R_V$.  Aguirre (1999) and Bianchi \& Ferrara (2005) argue that small dust grains are preferentially destroyed during the processes that expel dust from galaxies.  If correct, this would imply higher than usual values of $R_V$.  Our results indicate that for values in the range $3.1 < R_V < 5.5$, the resulting optical depths are nearly identical at low redshifts and slowly grow with increasing redshift.  For $\mathcal{Z}=\zsun$ out to $z=15$ at $\lambda_\mathrm{obs}=2~\mu\mathrm{m}$, $\tau_d=139$ and 94 for $R_V=3.1$ and 5.5, respectively. 

Because the DGR decreases with decreasing metallicity, the derived optical depths in Figure 1b are lower for an IGM metallicity of $0.13\zsun$, at which for $\lambda_\mathrm{obs}=2~\mu\mathrm{m}$, $\tau_d$ reaches only a modest value of 3.6 by $z=15$.

\subsection{Comparison with Observations}
Figure 1 also shows extinction measurements of the IGM drawn from the literature.  M\'enard et al. (2010b) analyzed the brightnesses of more than 20,000 quasars to derive optical extinction of 24 million foreground SDSS galaxies; they derived an average optical extinction of $\av = 0.03$ mag out to a redshift of $z=0.5$. Johansson \& M\"ortsell (2012) used quasar colors to estimate an upper limit of $\av\lesssim 0.25$ mag out to $z=1$.  Using $A(\lambda)\approx 1.086\tau_d(\lambda)$ (e.g., Draine 2011), we convert these values to optical depths and label them with diamond and star symbols in Figure 1.  

For higher redshifts, we overplot data points from Gallerani et al. (2010) and Bouwens et al. (2015).  Gallerani et al. (2010) inferred the amount of extinction toward $3.9<z<6.4$ quasars by fitting the quasar spectra with a variety of extinction curve templates.  They found that 7 of 33 quasars required extinctions at 3000 \AA~of $0.8\le A_{3000}\le 2$ mag.  In their examination of the UV luminosity function, Bouwens et al. (2015) analyzed more than 10,000 galaxies in the redshift range $4\le z\le 10$.  To quantify the amount of extinction by dust, they used a relation between the spectral slope $\beta$ and rest frame extinction at $\lambda=1600$ \AA~derived  by Meurer et al. (1999).   The relation, $A_{1600} = 4.91+2.21\beta$, originates from measuring the deviation of the spectral slope from an average slope representing no reddening, $\beta\approx -2.2$, (close to the Rayleigh-Jeans tail of a blackbody for which $\beta=-2$).  Note that a rest wavelength of 3000 \AA~at $z=6.4$, the highest redshift probed by Gallerani et al. (2010), corresponds to an observed wavelength of $2.2~\mu\mathrm{m}$; a rest wavelength of 1600 \AA~at $z=10.4$, the highest redshift in the Bouwens et al. (2015) study, corresponds to $\lambda_\mathrm{obs}=1.8~\mu\mathrm{m}$.

M\'enard et al. (2010b) and Johansson \& M\"ortsell (2012) considered dimming due to intergalactic dust in the rest frame $V$- and $B$-bands, at $z=0.5$ and $z=1$, respectively, corresponding to $\lambda_\mathrm{obs}\sim 1~\mu\mathrm{m}$.  In Figure 1a, the data points representing the results of these two studies fall just below the $\tau_d$-$z$ relation for $\lambda_\mathrm{obs}=1~\mu\mathrm{m}$ (red line), whereas the data lay well above the predicted relation for lower metallicities (Figure 1b).  This suggests that at low redshifts, a highly enriched IGM is plausible.  Indeed, the metallicity inferred by both observations and simulations for the intercluster medium of galaxies is $\sim 0.3\zsun$ (e.g., De Grandi et al. 2004; Nagashima et al. 2005; Elkholy et al. 2015).  Even if dust may be destroyed in the hot intracluster medium, it may survive in the cold IGM.

\section{Discussion and Conclusions}
The $\tau_d$-$z$ relation at $\lambda_\mathrm{obs}=2~\mu\mathrm{m}$ for $\mathcal{Z}=0.13\zsun$ saturates the highest redshift data point at $z\sim 10$ in Figure 1b, but most of the associated dimming along the line of sight likely originates from the ISM of the source galaxies. 
Thus, we limit the metallicity of the IGM to $\mathcal{Z}\lesssim 0.13\zsun$ by $z=10$.  Equation (\ref{eq:dgr}) leads to \emph{an upper limit on the intergalactic DGR of} $1.9\times 10^{-4}\approx 0.03(\mathrm{DGR}_\odot)$.

Figure 2a plots upper limits of the DGR as a function of redshift.  Each upper limit on the DGR is derived by calculating the metallicity at which the $\tau_d - z$ relation passes through the measured optical depth, for a given $\lambda_\mathrm{obs}$.  The envelope of our limits at all redshifts, combined with the detection of intergalactic dust at $z\lesssim 1$ (M\'enard et al. 2010), imply that the intergalactic DGR evolves with redshift.  That the DGR evolves with redshift is further emphasized in Figure 2b, which shows the upper limit envelopes of the DGRs derived for all sources.  For a given source at redshift $z_s$, the optical depth to any redshift $z^\prime<z_s$ must be less than the optical depth at the source.  For instance, at $z_s = 10.4$, the optical depth limit is 1.29, and we may derive the DGR for all $z^\prime<z_s$, assuming that $\tau_d(z^\prime)<1.29$.  The curves shown in Figure 2b, therefore, provide further constraints on the DGR for redshifts where observations do not exist.

Schaye et al. (2003) measured the statistical correlation between \ion{C}{4} and \HI~absorption in a sample of quasars in the redshift range $1.8<z<4.1$, to find that metallicity evolves only weakly, if at all, during this era.  For their representative model, in terms of the carbon-to-hydrogen abundance, they find $(\mathrm{C/H})\approx 1.6\times 10^{-3}(\mathrm{C/H})_\odot$.  Their results are consistent with Songaila (2001) who determined from the \ion{C}{4} distribution towards 32 quasars that the cosmic metallicity is greater than $3.5\times 10^{-4}$ by $z=5$.  Figure 2 yields a weaker upper limit on the metallicity of $\sim 0.28\zsun$ at $z=4$, about 156 times the Schaye et al. result.  We note that Schaye et al.'s  analysis does not uncover carbon in hot, X-ray emitting gas, nor does it reveal neutral or singly ionized carbon concealed in cold, self-shielded gas.  Thus, their results should be regarded as lower limits on the carbon abundance.  Moreover, the Schaye et al. (2003) and Songaila (2001) studies do not consider the \emph{total} metal abundance of the IGM, rather they focus on select elements.

Recently, Watson et al. (2015) reported on the dust properties of a $z=7.5$ galaxy---observed with the Atacama Large Millimeter Array (ALMA)---representative of a star-forming system during the epoch of reionization.  They infer $\t{DGR}=1.7\times 10^{-2}$, a few times above the Milky Way value.  This is somewhat surprising, considering previous expectations that early galaxies may be similar to local dwarf galaxies, which are dust- and metal-poor (e.g., Fisher et al. 2014).  What we do \emph{not} yet know about this young galaxy, A1689-zD1, is the degree of reddening it has experienced due to its enormous dust content.  Nor do we know the dust composition or size distributions, both of which would factor into measurements of the dust optical depth.  This compelling result by Watson et al. (2015), and the other interesting findings that are sure to follow from ALMA, only raise more questions about the nature of dust and star formation in the early Universe.  

\acknowledgments
We thank Anastasia Fialkov, Pascal Oesch, and Steven Furlanetto for their helpful comments on the paper.  This work was supported in part by NSF grant AST-1312034.


\end{document}